\documentclass[floatfix,letterpaper,longbibliography,nofootinbib,aps,prl,reprint,superscriptaddress,showkeys,showpacs,10pt]{revtex4-1}
\usepackage{caption}
\usepackage[caption=false]{subfig}
\usepackage{amsmath, amsthm, amssymb}
\usepackage{graphicx}
\usepackage{float}
\usepackage{bm}                                    
\usepackage{url}
\usepackage{placeins}
\newcommand\numberthis{\addtocounter{equation}{1}\tag{\theequation}}

\numberwithin{equation}{section}
\begin{document}
\title{Perfectly Reflectionless Omnidirectional Absorbers and Electromagnetic Horizons}
\date{\today}
\author{Kamalesh Sainath}
\email{sainath.1@osu.edu}
\author{Fernando L. Teixeira}
\email{teixeira@ece.osu.edu}
\affiliation{The Ohio State University: ElectroScience Laboratory}
\altaffiliation[Address: ]{1330 Kinnear Road, Columbus, Ohio, USA 43212}
\begin{abstract}
\noindent
We demonstrate the existence of metamaterial blueprints describing, and fundamental limitations concerning, perfectly reflectionless omnidirectional electromagnetic absorbers (PR-OEMA). Previous attempts to define PR-OEMA blueprints have led to active (gain), rather than passive, media. We explain this fact and unveil new, distinct limitations of true PR-OEMA devices, including the appearance of an ``electromagnetic horizon'' on physical solutions. As practical alternatives we introduce alternative OEMA blueprints corresponding to media that, while not reflectionless, are nonetheless effective in absorbing incident waves in a manner robust to incident wave diversity.
\end{abstract}

\maketitle
\section{Introduction}

The concept of a polarization-independent, perfectly reflectionless omnidirectional electromagnetic absorber (PR-OEMA) has its origins in ``Perfectly Matched Layer" (PML) media~\cite{berenger}. These PML media, in the continuous-medium limit, possess matched electric and magnetic biaxial anisotropies that preferentially attenuate electromagnetic (EM) waves, \emph{without reflection}, along a particular direction of propagation~\cite{jflee,teixeiraJEWA,chew4,teixeira21,ref4}.\footnote{Some notes are in order. First, we refer to reflectionless absorption within a \emph{particular} frequency band, which typically implies \emph{incoherent} emission at another band. Second, it is implicitly understood that reflectionless behavior in general only exists in the continuous-medium limit. Discretizations of the material profile, whether in simulations or physical realization, may induce reflections~\cite{chew11}. Third, references to ``concave" or ``convex" surfaces are with respect to the source location (or incident field). Fourth, when defining material tensor parameters through equivalent coordinate transformations, our definitions of the flat-space and ``deformed" (``complex-space") coordinates follow much of the PML literature~\cite{teixeiraJEWA}, and are reversed with respect to the Transformation Optics literature (the final PML material properties are, of course, independent of coordinate convention)~\cite{pendry10}. Fifth, without loss of generality we assume herein that the ambient medium, into which we embed the OEMA devices, is vacuum. Extension to more general ambient media not supporting ``backward-wave" modes follows~\cite{becache,sjohnson2,pendry1}.} Their absorptive and reflectionless properties have already been widely exploited in computational physics
for over two decades~\cite{chew6,teixeiraJEWA,jflee,berenger,becache,hu2,lantos,leonhardt2,chew8,ref4}.

Furthermore, there is great interest in {\it convex}, geometrically-conformal coatings to absorb incident\footnote{This is in contrast to the use of \emph{concave} or planar geometrically-conformal absorbers used in most computational simulations, which are placed on the external boundary of the computational domain to absorb \emph{outgoing} waves~\cite{berenger}.} EM radiation upon finite objects. Indeed many theoretical, numerical, and experimental efforts have already been made concerning minimally-reflective, wide-angle absorbers (e.g., see~\cite{soukoulis5,narimanov1,argyr,cui1}). However, in comparison to isotropic-medium absorbers~\cite{narimanov1}, absorbers possessing electric and magnetic biaxial anisotropy can exhibit performance more robust to incident signal diversity (important, for example, when designing high-performance radar and optical absorbers) with respect to wave polarization, incidence angle spectrum, and temporal frequency spectrum, while being both thin and minimally-scattering~\cite{berenger,jflee,teixeira12,teixeira11,teixeira16,petro}.

Previous results have indicated that PML absorber blueprints can be derived and used for coating planar or concave surfaces, but not so for \emph{convex} surfaces~\cite{teixeiraJEWA,chew6,teixeira16,teixeira11}. While our results here corroborate these conclusions in a broader sense, there are two compelling reasons for revisiting this question. The first reason is to comprehensively examine and explain the fundamental roadblocks on all four canonical design pathways (discussed below) whereby one may attempt to prescribe convex-conformal PR-OEMA devices. Interestingly, we show that blueprints for ``primitively'' causal (i.e., causes preceding effects) PR-OEMA devices can, in a certain sense, be defined; however, they too ``fail" due to the absorptive core being \emph{necessarily} cloaked from any external radiation by an ``EM horizon'' to satisfy energy conservation. The second reason is that revisiting this question suggests \emph{ad hoc} modifications of convex-conformal PR-OEMA devices which we show, through full-wave numerical simulations in Section 3, to yield excellent omnidirectional, broadband, and polarization-robust absorption performance with extremely low (albeit non-zero) backscatter. Throughout this paper, we use the convention exp($-i \omega t$) for time-harmonic fields.

\section{Fundamental Roadblocks for PML Media Conforming to Convex Surfaces}
\label{form}

Consider a filament of current, located in vacuum near a perfect electric conductor (PEC) cylinder of radius $\rho_2$; a PML is inserted in the region $\rho_2\leq \rho \leq \rho_0$ coating the PEC cylinder. Recall that a basic principle guiding the design of the PML's material properties is to obviate \emph{any} scattered fields; mathematically speaking, this means that the boundary conditions within the region $\rho \leq \rho_0$ must be left unperturbed~\cite{teixeiraJEWA}. In the frequency domain, the EM fields within the cylindrical PML ($\rho_2 \leq \rho \leq \rho_0$) can be mathematically derived through a very simple transformation of the radial variable $\rho$ (``complex-space" PML interpretation)~\cite{teixeiraJEWA,chew4}:
\begin{equation}
\bar{\rho}(\rho,\omega)=\rho_0+\int\limits_{\rho_0}^{\rho}s_{\rho}(\rho',\omega)\mathrm{d}\rho', \ \frac{\mathrm{d}\bar{\rho}}{\mathrm{d}\rho}=s_{\rho}(\rho,\omega) \label{e0}
\end{equation}
where
$s_{\rho}(\rho,\omega)$ is the so-called ``coordinate stretching'' variable. Under the hypothesis that $s_{\rho}(\rho,\omega)$ is a bounded function of the spatial argument, it immediately follows that that the transverse electric and magnetic field components are \emph{continuously} transformed from the original vacuum fields (e.g., $\mathcal{E}_{z}(\rho,\phi,z)$) to the new fields (e.g., $\mathcal{E}_{z}(\bar{\rho}[\rho],\phi,z)$) within the PML. From Transformation Optics (T.O.) principles~\cite{teixeiraJEWA,pendry6}, a PML interpretation equivalent to the ``complex-space" one can be obtained whereby the PML is represented instead by a (doubly) anisotropic medium, with relative electric permittivity tensor $\boldsymbol{\bar{\epsilon}}_r$ and relative magnetic permeability tensor $\boldsymbol{\bar{\mu}}_r$ given by
\begin{align}
\epsilon_{\rho\rho}&=\mu_{\rho\rho}=T_{\rho\rho}=\bar{\rho}/(\rho s_{\rho}) \numberthis \label{e1}\\
\epsilon_{\phi\phi}&=\mu_{\phi\phi}=T_{\phi\phi}=T_{\rho\rho}^{-1} \numberthis \label{e2} \\
\epsilon_{zz}&=\mu_{zz}=T_{zz}=s_{\rho}\bar{\rho}/\rho \numberthis \label{e3}
\end{align}
Up to this point we implicitly assumed a reflectionless, \emph{lossless} PML medium, and indeed Eqns. \eqref{e1}-\eqref{e3} would result in an inserted medium (e.g., an EM cloak~\cite{pendry10}) that leaves unperturbed the fields within the vacuum region.

How does one include absorption as well? In the frequency domain, the answer (in principle, at least) is also quite simple: make $s_{\rho}$ complex-valued in Eqns. \eqref{e0}-\eqref{e3}~\cite{teixeiraJEWA}. Indeed, the imaginary part of $s_{\rho}$ (Im[$s_{\rho}$]) controls the wave's absorption or amplification~\cite{teixeira12}. One dispersive, causal form for $s_{\rho}$ and $\bar{\rho}$, originating from the low-frequency limit of the Drude dispersion model,\footnote{Other dispersion models are also possible~\cite{cummer}.} is~\cite{jflee,teixeira12}:
\begin{equation}
s_{\rho}(\rho,\omega)=a(\rho)+\frac{i\sigma_0(\rho)}{\omega}, \
\bar{\rho}(\rho,\omega)=b(\rho)+\frac{i\Delta_0(\rho)}{\omega} \label{e4}
\end{equation}

Despite the ease with which we can derive these complex-valued material parameters, the real challenge is how to simultaneously satisfy the constraints of passivity (and primitive causality) while retaining the device's reflectionless and geometrically conformal characteristics.
Let us start exploring this issue by setting Re[$s_{\rho}] > 0$ and Im[$s_{\rho}]\geq 0$ (the first of the four canonical PML design strategies examined in this Section), which leads to passive absorbers when coating \emph{concave} geometries but instead leads to active (gain) media when coating \emph{convex} geometries~\cite{teixeira16}. The behavior of cylindrical PML media arising from this design procedure is illustrated in Figs. 1a-1b, which show that the outer, concave surface-conforming PML is able to absorb EM waves without reflection (up to numerical discretization artifacts). On the other hand, the inner PML at first appears to be absorbing incident waves as they enter the device from the right (see Fig. 1a). However, as the waves propagate further into the device, the active medium behavior becomes apparent (see Fig. 1b).

 \begin{figure}[H]
\centering
\fbox{\includegraphics[width=3.24in,height=2.88in]{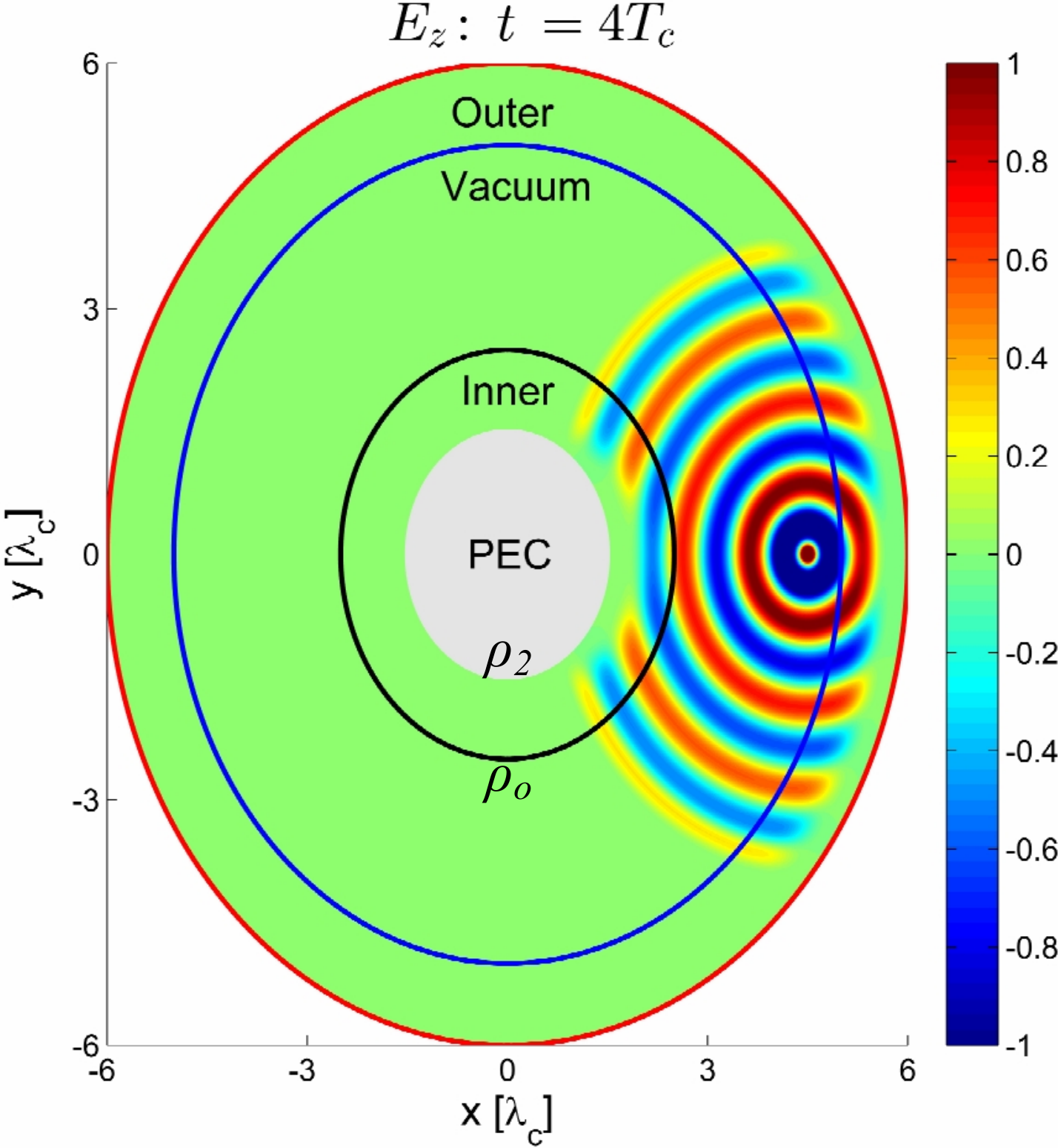}}

\fbox{\includegraphics[width=3.24in,height=2.88in]{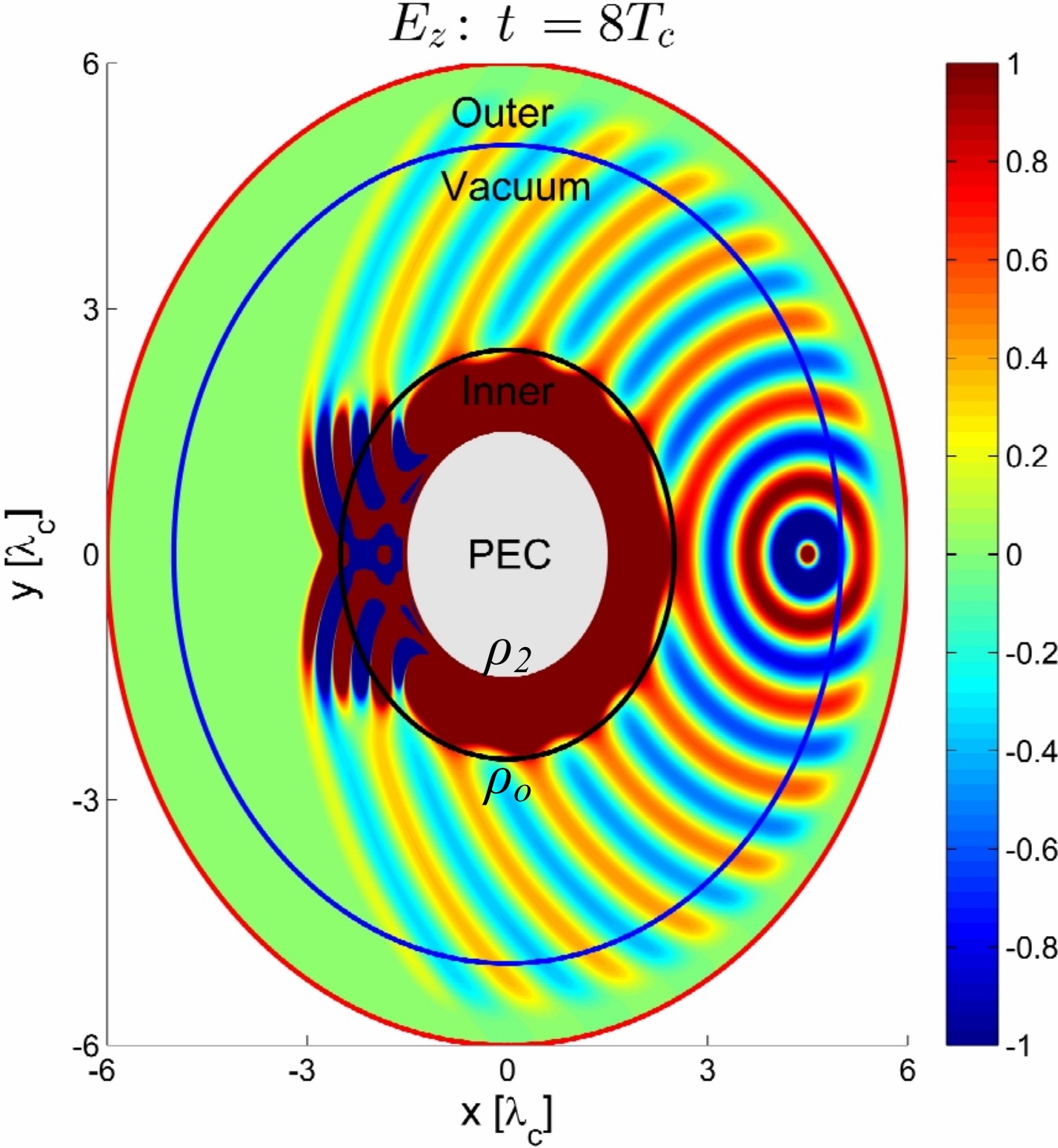}}
\caption{(Color online) Electric current radiating a ramped-sinusoidal wave (sinusoid frequency $f_c=1/ T_{c}=c/ \lambda_c$) in vacuum near a PEC cylinder ($ \rho_2 = 1.5 \lambda_c $) coated by a $1 \lambda_c$ thick convex PML (``Inner"); $\rho_0=2.5 \lambda_c$. The outer PEC wall (red line) is also coated by a concave PML (``Outer") simply to mimic open space. Both PMLs are designed using Re[$s_{\rho}] > 0$ and Im[$s_{\rho}]\geq 0$. Note the asymmetry on the field behavior, with the concave PML behaving as an absorber but the inner PML behaving as a gain (active) medium.}
\end{figure}	

\begin{figure}[H]
\centering
\fbox{\includegraphics[width=3.24in,height=2.88in]{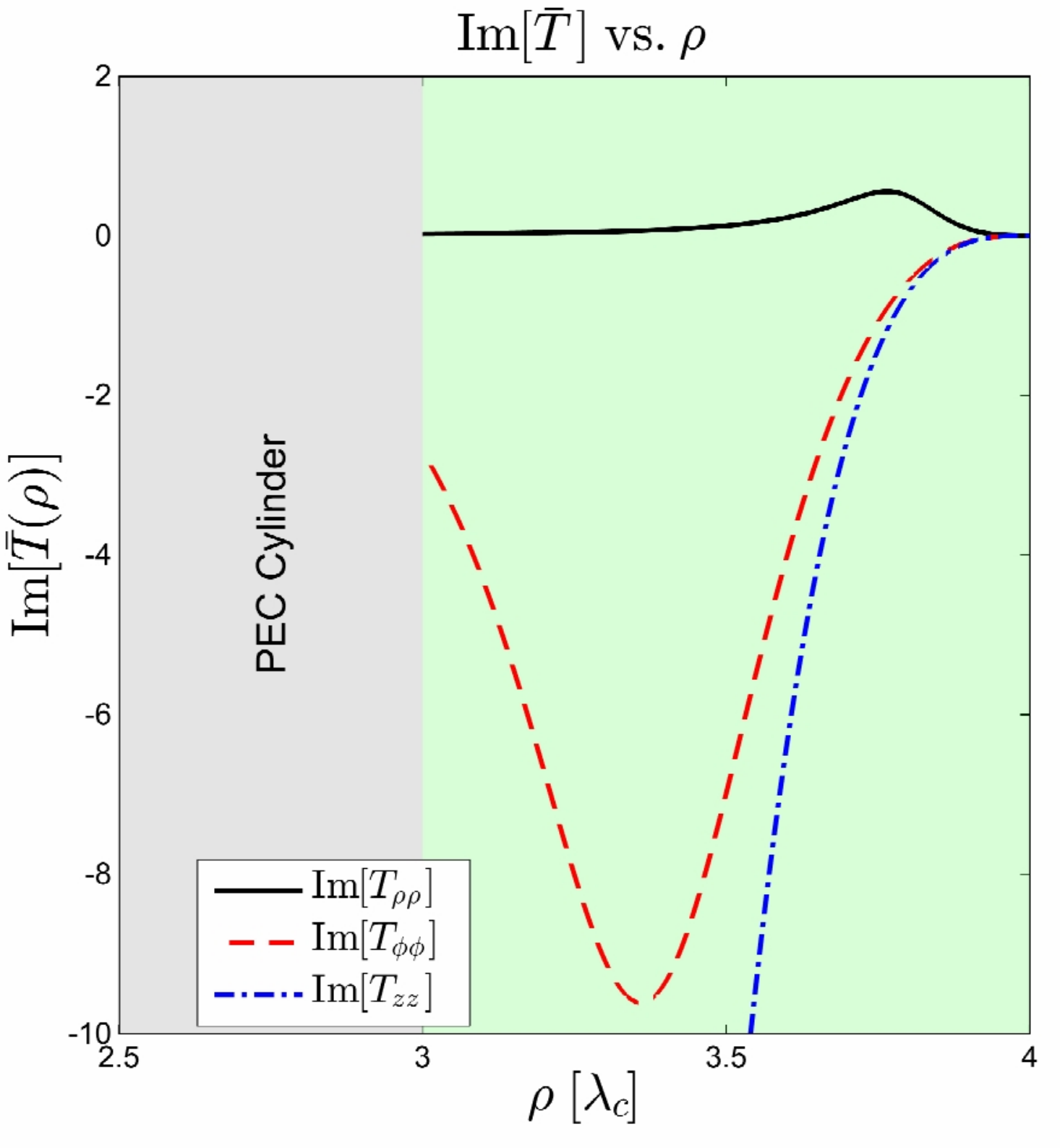}}

\fbox{\includegraphics[width=3.24in,height=2.88in]{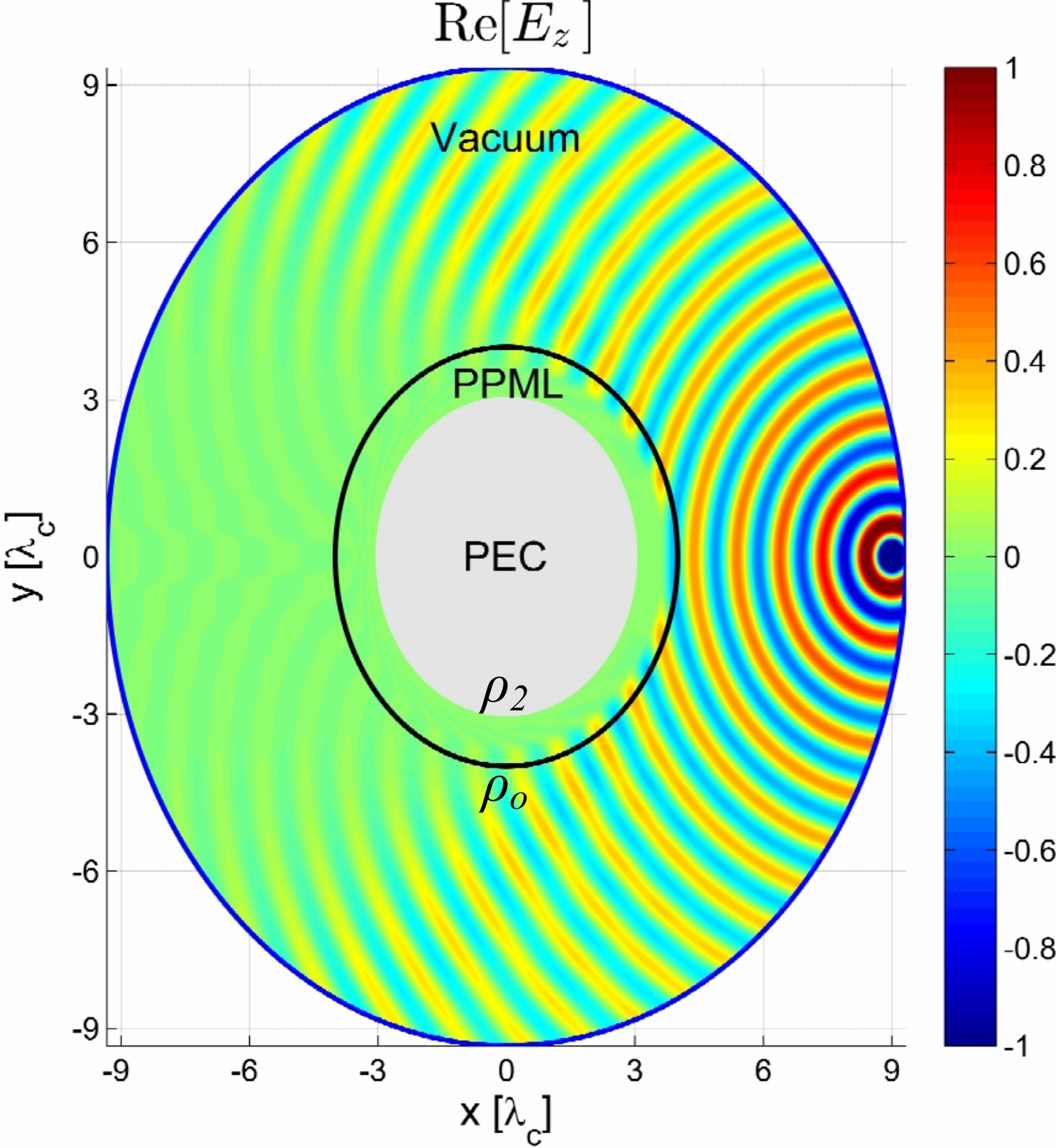}}
\caption{\small (Color online) Fig 2a: Imaginary part of the material tensor components of a convex-conformal (gain) PML medium designed using the fourth canonical PML design strategy, viz. Im[$s_{\rho}]\leq 0$ and Re[$s_{\rho}] < 0$. Fig 2b: Frequency-domain Finite Element Method simulation (COMSOL RF Module) demonstrating the absorption and scattering performance of the ``pseudo" version of the PML described in Fig. 2a ($\rho_2=3\lambda_c,\rho_0=4\lambda_c$). The ``pseudo" PML (PPML) material blueprint (described in Section 3) is prescribed using Eqn. \eqref{e0} followed by evaluating Eqns. \eqref{e1}-\eqref{e3} with an \emph{ad hoc} sign-reversed version of $s_{\rho}$ within the PML's absorptive region (i.e., Im$[s_{\rho}]\neq 0$).}
\end{figure}	
Intuitively, why does the convex PML exhibit gain behavior? Well, we designed a perfectly reflectionless absorber of EM waves within a  spatially \emph{localized} (finite) region: If the EM field is to comply with energy conservation, then to both absorb waves within a finite-sized region while also leaving the waves unperturbed outside this region demands the presence of an energy ``pump" (within said region) to restore back the energy (c.f. Fig. 1b) that was initially absorbed (c.f. Fig. 1a). Interestingly enough, however, as has been observed elsewhere~\cite{teixeira12} the energy pump has the added effect of amplifying the initially-absorbed waves in such a way so as to corrupt (saturate) the wave solution outside the device; this can be observed in Fig. 1b just to the left of the solid black line, which demarcates the interface between vacuum and the inner (convex) PML. As an aside, we mention that this behavior can be predicted~\cite{teixeira16} from the analytical properties, in the complex $\omega$ plane, of the PML absorber constitutive tensors $\boldsymbol{\bar{\epsilon}}_r$ and $\boldsymbol{\bar{\mu}}_r$. The same $\omega$-plane analysis also dooms to ``failure" two of the three remaining possible canonical design options with Re[$s_{\rho}] > 0$ and Im[$s_{\rho}]\leq 0$, or with Re[$s_{\rho}] < 0$ and Im[$s_{\rho}]\geq 0$ ~\cite{teixeira16}, which we do not explore further.

However, these considerations do not apply for the fourth canonical design pathway, Im[$s_{\rho}]\leq 0$ and Re[$s_{\rho}] < 0$.
In this case, (i) Re[$s_{\rho}]$ and Im[$s_{\rho}]$ share the same (negative) sign while, for a convex surface-conforming PML, (ii) Re[$\bar{\rho}] > 0$ and Im[$\bar{\rho}]\geq 0$ would also share the same (positive) sign. For our example of a Drude-based, broadband PML medium, said $\omega$-plane analysis reveals that these two conditions are necessary and sufficient for a primitively causal medium~\cite{teixeira16,teixeira12}. However, such a ``double-negative" medium PML \emph{also} would, in fact, amplify transverse EM waves but for a different reason: the radial component of the group velocity (energy flow) and wave vectors become then anti-parallel; see a detailed discussion of the associated amplification mechanism in~\cite{cummer,becache,sjohnson2}. Active medium behavior can also be anticipated from the PML material profile arising from this design pathway, as illustrated in Fig. 2a; indeed $T_{zz}$ (=$\epsilon_{zz}$=$\mu_{zz}$) has negative imaginary part, which corresponds to \emph{active} behavior~\cite{chew}[Ch. 2]. Hence a transverse-magnetic-to-$z$ (TM$_z$) wave ($\mathcal{E}_z$) will ``sense" the active medium behavior induced by $\epsilon_{zz}$, and similarly for a transverse-electric-to-$z$ (TE$_z$) wave ($\mathcal{H}_z$) due to sensing the active behavior induced by $\mu_{zz}$.

\section{The PR-OEMA, Practical Alternatives, and Numerical Results}

\subsection{PR-OEMA and EM Horizon Emergence}

We propose a ``non-canonical'' blueprint for the PR-OEMA by embedding an annular ``core" ($a(\rho)< 0$, $\sigma_0(\rho) \leq 0$), occupying the inner PR-OEMA region $\rho_2 \leq \rho \leq \rho_1$, \emph{inside} a non-absorptive outer region ($a(\rho) > 0$, $\sigma_0(\rho)=0$) occupying $\rho_1 \leq \rho \leq \rho_0$. In particular, the non-absorptive region's role is to transform $\bar{\rho}$ (the \emph{effective} spatial radius within the PR-OEMA) from its free-space value $\bar{\rho}=\rho$ at $\rho=\rho_0$ to $\bar{\rho}(\rho)=0$ at $\rho=\rho_1$ (the physical interpretation of, and reasoning behind, this radial transform will be apparent in a moment). To avoid active-medium behavior, after prescribing the reflectionless (but \emph{active}) inner region's material parameters, the sign of $s_{\rho}$ within said inner region is reversed in an \emph{ad hoc} fashion to recover absorptive behavior. One then uses the true PR-OEMA's $\bar{\rho}$ and the pseudo $s_{\rho}$ to prescribe the material blueprint via Eqns. \eqref{e1}-\eqref{e3}.

We remark that other types of \emph{ad hoc} modifications in the coordinate transformation have been prescribed before to restore passive medium behavior at the expense of sacrificing the reflectionless property, such as using the design pathway Re[$s_{\rho}] > 0$ and Im[$s_{\rho}]\geq 0$ (leading to Re[$\bar{\rho}]> 0$, Im[$\bar{\rho}]\leq 0$) followed by an \emph{ad hoc} sign reversal of Im[$\bar{\rho}]$ to restore passive (albeit also reflective) behavior~\cite{teixeira12}. What makes the ``pseudo" PR-OEMA proposed herein unique is that, under a particular, additional condition\footnote{That is, a condition beyond the non-absorptive, outer PR-OEMA region itself being perfectly impedance matched both internally (i.e., inside said outer region) and to vacuum at $\rho=\rho_0$.} satisfied by the outer PR-OEMA region, the PR-OEMA as a whole \emph{is} in fact perfectly reflectionless and leaves unperturbed the external ($\rho\geq \rho_0$) fields. This can be succinctly shown as follows. First, one can write the wave equation for a time-harmonic TM$_z$ wave within the initially-defined (gain) PR-OEMA \emph{and} ``pseudo" (passive) PR-OEMA\footnote{Indeed, an \emph{ad hoc} sign reversal of $s_{\rho}$ leaves unchanged the wave equation Eqn. \eqref{waveeq} satisfied by fields in the passive ``pseudo" PR-OEMA.} ($k_0=2\pi / \lambda_0$, $\nu$ is the azimuth index)~\cite{petro}:
\begin{equation}
\frac{\bar{\rho}}{s_{\rho}}\frac{\partial}{\partial \rho}\left(\frac{\bar{\rho}}{s_{\rho}}\frac{\partial }{\partial \rho}\mathcal{E}_z \right)+(k_0^2\bar{\rho}^2-\nu^2) \mathcal{E}_z =0 \label{waveeq}
\end{equation}
with standard cylinder wave solutions $\{J_{\nu}(k_0\bar{\rho}),Y_{\nu}(k_0\bar{\rho}),H_{\nu}^{(1)}(k_0\bar{\rho}),H_{\nu}^{(2)}(k_0\bar{\rho})\}\mathrm{exp}(i\nu \phi)$. Second, consider the relation between $\mathcal{E}_z$ and $\mathcal{H}_{\phi}$ within the inner region of the true (gain) and ``pseudo" (passive) PR-OEMA ($\zeta=k_0\bar{\rho}$)~\cite{petro}
\begin{equation}
\mathcal{H}_{\phi} = \frac{\pm i\omega \epsilon_0 \bar{\rho}}{k_0\rho}\frac{\partial\mathcal{E}_z}{\partial \zeta} \label{hphi},
\end{equation}
where one takes (resp.) the top or bottom sign for the true (gain) PR-OEMA or the \emph{ad hoc}-modified (passive) ``pseudo" PR-OEMA.\footnote{Note that the sign reversal in the ``pseudo" PR-OEMA's Eqn. \eqref{hphi}, versus the true PR-OEMA case, is due to the true PR-OEMA's $s_{\rho}=\mathrm{d}\bar{\rho}/\mathrm{d}\rho$ having the opposite sign as the ``pseudo" (sign-reversed) $s_{\rho}$~\cite{petro}. Also note that within the outer non-absorptive region where no \emph{ad hoc} sign reversal in $s_{\rho}$ occurs (and hence where the two OEMAs share identical material properties), one takes the plus sign in Eqn. \eqref{hphi} for both OEMAs.} A similar conclusion can also be established for TE$_z$ waves. By writing down the expression for the 2$\times$2 Fresnel reflection matrix associated with an incident field (in general a combination of TE$_z$ and TM$_z$ polarizations) at the boundary of two cylindrical layers~\cite{chew}[Ch. 3], one can show that zero reflection results at the boundary $\rho=\rho_1$ of the PR-OEMA's outer and inner regions if $\bar{\rho}(\rho_1)=0$. Slightly more involved derivations, starting from Eqn. \eqref{waveeq}, can be repeated to obtain similar conclusions for conical waves possessing non-zero longitudinal wave number component $k_z$~\cite{petro}.


From a fundamental perspective, the coordinate transformation defining the outer PR-OEMA region's material parameters reveals that said outer region acts as an ``EM horizon'' that ``cloaks" the absorptive inner core~\cite{pendry6}. The presence of this horizon precludes the absorptive core's interaction with external radiation, and hence renders the designed PR-OEMA useless from a practical standpoint. Note that, given the incompatibility between reflectionless and omnidirectional absorptive properties when referred to such an interior (i.e., spatially \emph{localized}) domain, the latter statement concerning the manifestation of an ``EM horizon" appears, in retrospect at least, inevitable if one is to satisfy energy conservation (as discussed in Section 2). In other words, if no internal energy restoration mechanism is present (c.f. Fig. 1), spatially localized wave absorption must (by \emph{necessity}) perturb the wave outside the absorber. Nevertheless, it is interesting to unveil here the precise mechanism whereby the presence of such an ``EM horizon'' at $\rho=\rho_1$ (i.e., where $\bar{\rho}(\rho)=0$), around which light travels~\cite{ref3} (akin to the inner boundary of other perfect invisibility cloaks; e.g., see~\cite{pendry10}), is \emph{forced} upon any convex-shaped, finite-sized object that attempts to absorb energy from the incident EM wave, leave the EM field distribution outside it completely unperturbed, yet also not return any of the absorbed energy. Indeed, the mathematical and physical \emph{necessity} of the cloaking represents the main result of the preceding PR-OEMA material derivation. This necessity stands in stark contrast to hypothetically prescribing the cloaking in some arbitrary, \emph{ad hoc} manner as a ``blanket" solution due to its capability of shielding any internal object(s) from interaction with externally incident waves~\cite{pendry10}. Next we discuss more practical alternatives for low-scattering absorbers which, while still challenging to physically realize with current technology, can at present be easily realized computationally to accomplish great time and resource savings in numerical simulations (for both frequency and time domain-based numerical methods) by allowing efficient domain truncation.

\begin{figure}[h]
\centering
\fbox{\includegraphics[width=3.24in,height=2.88in]{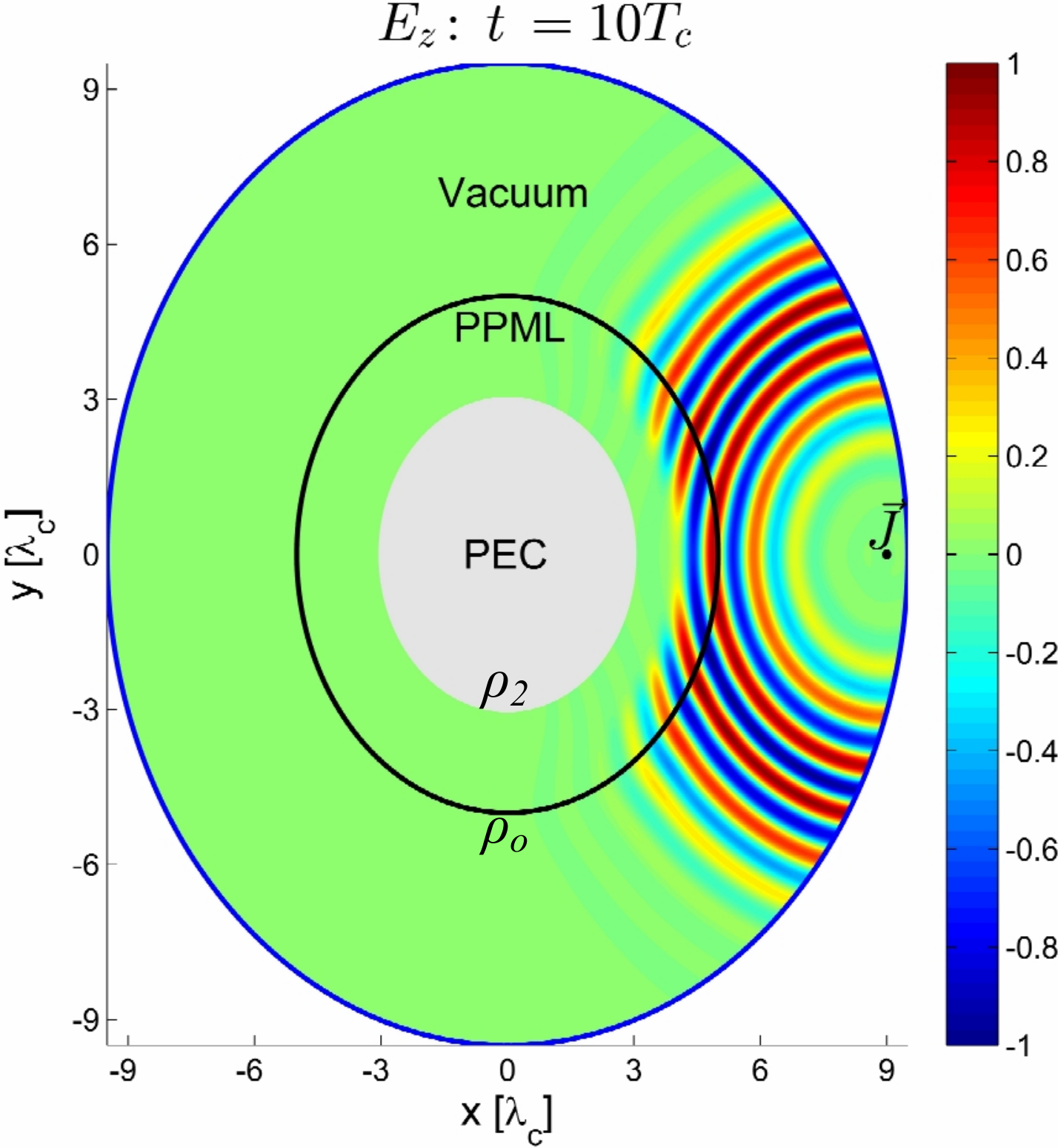}}

\fbox{\includegraphics[width=3.24in,height=2.88in]{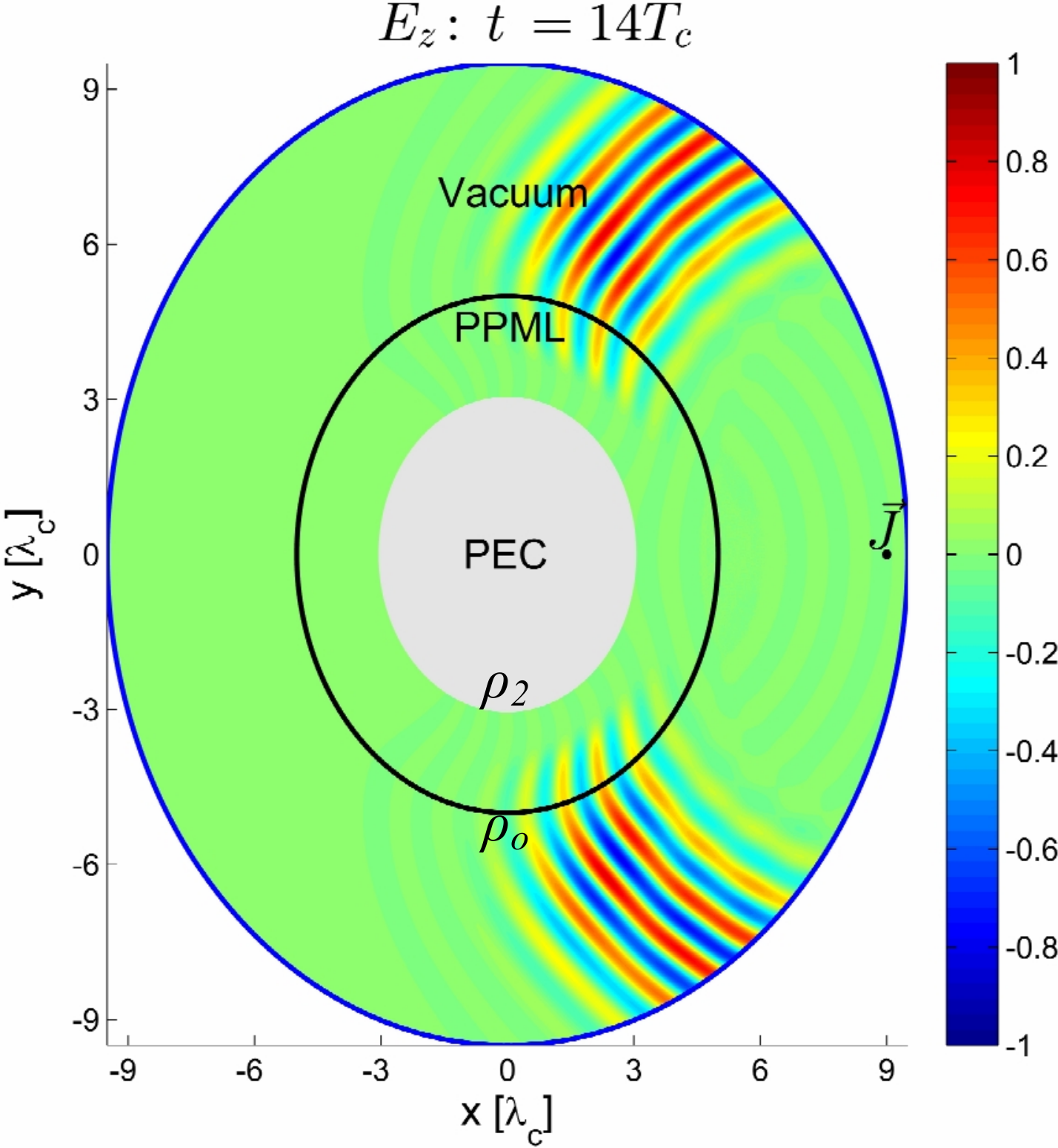}}
\caption{\small (Color online) ``Pseudo" PML (PPML) performance under TM$_z$ wave illumination. Time snapshots: $t=10T_c$ and $t=14T_c$.}
\end{figure}	

\subsection{Possible Practical Alternatives}

In light of said limiting ``EM horizon" characteristic of PR-OEMA structures, let us modify the material properties shown in Fig. 2a by reversing, again in an \emph{ad hoc} fashion, the sign of $s_{\rho}$ to restore absorptive behavior (although now, the sign reversal is performed throughout the \emph{entire} OEMA). Fig. 2b shows the frequency-domain simulation result of this modified OEMA's performance, where there is hardly any noticeable backscatter.\footnote{The absorber thickness $L$ has purposefully been designed as relatively thick ($L=1\lambda_c$) for visualization purposes only; a thinner absorber can be obtained through geometrical re-scaling~\cite{sjohnson}.} Such absorbers, based on (for example) the low-frequency limit of the Drude dispersion model, can also be broadband to facilitate efficient internal domain truncation via absorption of pulsed waves (e.g., in time-domain simulations). In Fig. 3, we again define $\bar{\rho}$ as per Eqn. \eqref{e0} followed by using this $\bar{\rho}$ along with an \emph{ad hoc} sign-reversed $s_{\rho}$ (again, done within the entire OEMA) to define the material parameters via Eqns. \eqref{e1}-\eqref{e3}. The incident signal is a Blackman-Harris (BH) pulse with central carrier frequency $f_c$ ($T_{c}=1/f_c=\lambda_c/c$) and $-10$dB field amplitude fractional bandwidth\footnote{$B=2\Delta f/f_c$, where $f_H=f_c+\Delta f$ and $f_L=f_c-\Delta f$ are the frequencies at which the magnitude of the BH window's Fourier spectrum equals 10\% ($-10$dB) of the spectrum magnitude's peak value at $f_c$.} $B$ of 46\%. As our cylindrical-grid finite-difference time-domain results show, the incident pulse is well absorbed with very low backscatter. Akin to the performance of other types of cylindrical PPML media however~\cite{tseng}, the subsections of the pulsed incident wavefront that (locally) are near grazing incidence to the cylindrical PPML-vacuum interface are not as effectively absorbed as compared to the wavefront subsections that (locally) are near normal incidence. Finally we remark that, from the symmetry in the material tensors (c.f. Eqns. \eqref{e1}-\eqref{e3}), PPMLs will perform very similarly for TE$_z$-polarized waves.


\section{Concluding Remarks}

We have discussed design strategies to prescribe PR-OEMA metamaterial blueprints that geometrically conform to convex cylindrical surfaces. After elucidating fundamental limitations of such strategies, we then proposed and numerically demonstrated an alternative, practical class of OEMA devices. While the proposed alternative OEMA devices are \emph{not} (by fundamental necessity) reflectionless, they exhibit very low backscatter with omnidirectional, broadband, and polarization-robust absorption capability.

\section{Acknowledgments}
This work was supported by the NASA-NSTRF program and by OSC under Grant PAS-0110. We acknowledge Mr. Z. Zeeshan and Mr. J. Burr of OSU-ESL for assistance in preparing COMSOL numerical results, as well as acknowledge Prof. Weng C. Chew for his input on an earlier version of the manuscript.

\begin{figure}[h]
\centering
\ContinuedFloat
\fbox{\includegraphics[width=3.24in,height=2.88in]{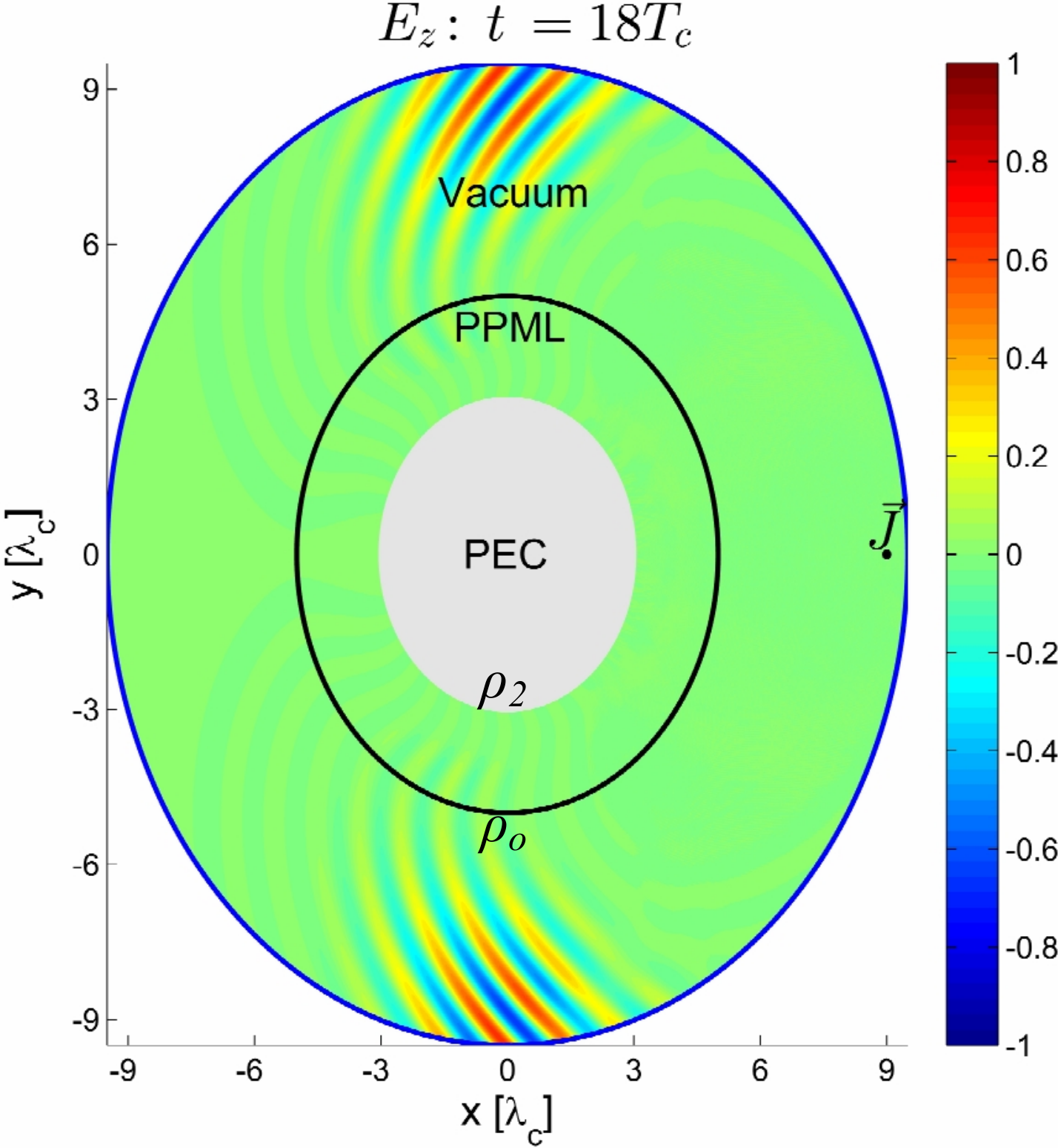}}

\fbox{\includegraphics[width=3.24in,height=2.88in]{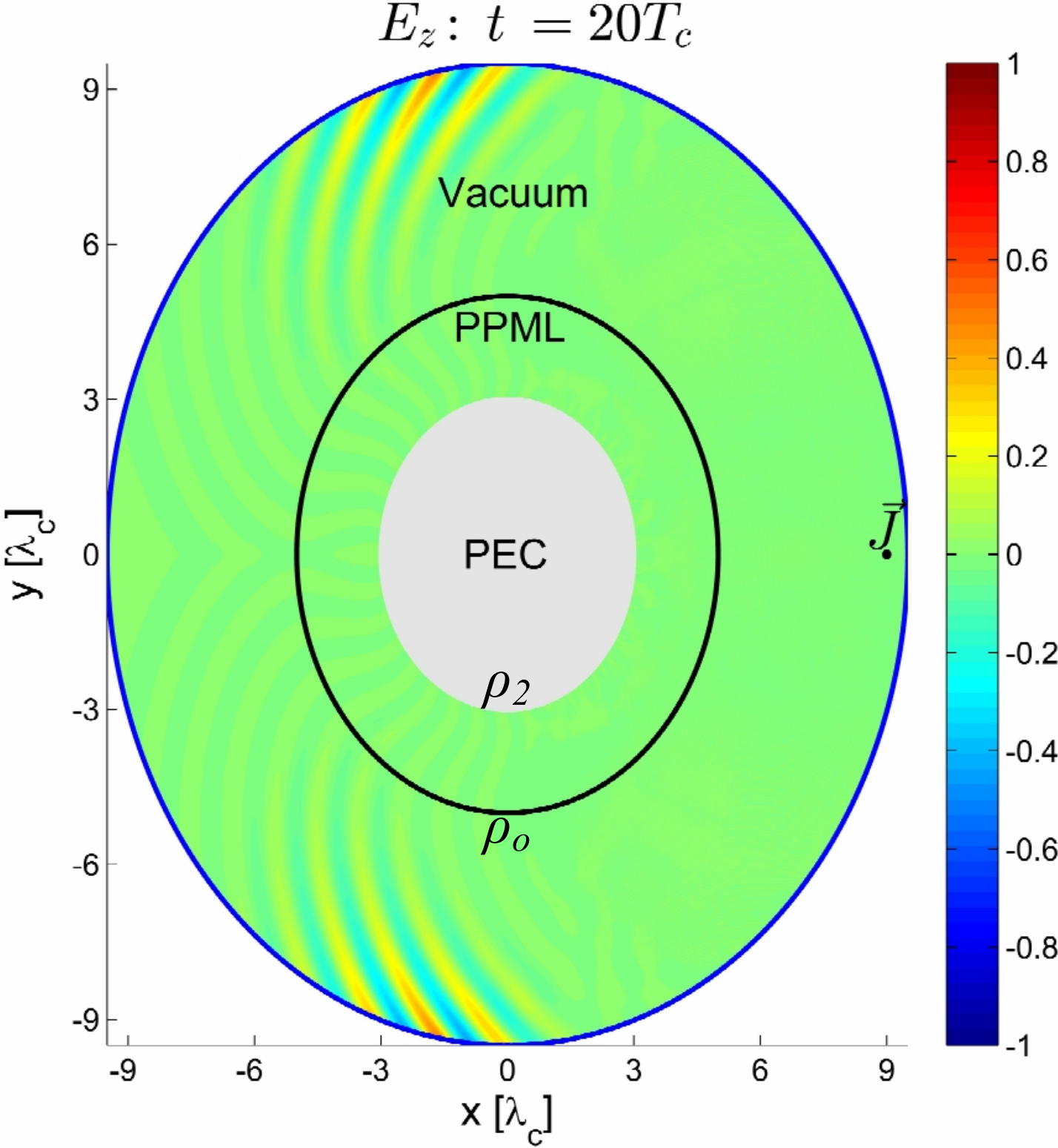}}
\caption{\small (Color online) PPML performance under TM$_z$ wave illumination (continued from prev. page). Time snapshots: $t=18T_c$ and $t=20T_c$.}
\end{figure}

\bigskip

\bibliography{reflist}
\bibliographystyle{apsrev4-1}
\end{document}